\documentclass[fleqn,usenatbib]{mnras}

\usepackage{newtxtext,newtxmath}

\usepackage[T1]{fontenc}
\usepackage{anyfontsize}
\usepackage{stackrel}
\usepackage{subfig}
\usepackage{bm}

\DeclareRobustCommand{\VAN}[3]{#2}
\let\VANthebibliography\thebibliography
\def\thebibliography{\DeclareRobustCommand{\VAN}[3]{##3}\VANthebibliography}

\usepackage{graphicx}	
\usepackage{amsmath}	

\newcommand{\tess}{\it TESS}

\newcommand{\chisq}{$\chi^{2}$}

\newcommand{\locccf}{CCF$_{\rm{loc}}$ }
\newcommand{\DIccf}{CCF$_{\rm{DI}}$ }
\newcommand{\locccfs}{CCFs$_{\rm{loc}}$ }
\newcommand{\DIccfs}{CCFs$_{\rm{DI}}$ }

\newcommand{\Nuonets}{\mbox{$0.358 \pm 0.014$}}
\newcommand{\Nutwots}{\mbox{$0.302 \pm 0.019$}}

\title[TOI-2119: spin-orbit alignment]{The First Spin-Orbit Obliquity of an M dwarf/brown dwarf System: An eccentric and aligned TOI-2119~b}

\author[L. Doyle et al.]{
Lauren Doyle,$^{1,2}$\thanks{E-mail: lauren.doyle@warwick.ac.uk}
Caleb I. Cañas,$^{3}$
Jessica E. Libby-Roberts,$^{4,5}$
Heather M. Cegla,$^{1,2}$
\newauthor
Guðmundur K. Stefánsson,$^{6}$
David Anderson,$^{7}$
David J. Armstrong,$^{1,2}$
Chad Bender,$^{8}$
Daniel Bayliss,$^{1,2}$
\newauthor
Theron W. Carmichael,$^{9}$
Sarah Casewell,$^{10}$
Shubham Kanodia,$^{11}$
Marina Lafarga,$^{1,2}$
Andrea S.J. Lin,$^{4,5}$
\newauthor
Suvrath Mahadevan,$^{4,5}$
Andy Monson,$^{8}$
Paul Robertson,$^{12}$
and Dimitri Veras$^{1,2}$
\\
$^{1}$Centre for Exoplanets and Habitability, University of Warwick, Coventry, CV4 7AL, UK \\
$^{2}$Department of Physics, University of Warwick, Coventry, CV4 7AL, UK \\
$^{3}$NASA Goddard Space Flight Center, Greenbelt, MD 20771, USA \\
$^{4}$Department of Astronomy \& Astrophysics, 525 Davey Laboratory, The Pennsylvania State University, University Park, PA 16802, USA \\
$^{5}$Center for Exoplanets and Habitable Worlds, 525 Davey Laboratory, The Pennsylvania State University, University Park, PA 16802, USA \\
$^{6}$Anton Pannekoek Institute for Astronomy, University of Amsterdam, Science Park 904, 1098 XH Amsterdam, The Netherlands \\
$^{7}$Instituto de Astronom\'ia, Universidad Cat\'olica del Norte, Angamos 0610, 1270709, Antofagasta, Chile \\
$^{8}$Steward Observatory, The University of Arizona, 933 N.\ Cherry Avenue, Tucson, AZ 85721, USA\\
$^{9}$Institute for Astronomy, University of Hawai‘i, 2680 Woodlawn Drive, Honolulu, HI, 96822, USA \\
$^{10}$Department of Physics and Astronomy, University of Leicester, University road, Leicester, LE1 7RH \\
$^{11}$Earth and Planets Laboratory, Carnegie Science, 5241 Broad Branch Road, NW, Washington, DC 20015, USA \\
$^{12}$Department of Physics \& Astronomy, The University of California, Irvine, Irvine, CA 92697, USA\\
}

\date{Accepted 2024 December 20. Received 2024 November 27; in original form 2024 September 27} 

\pubyear{2024}

\begin{document}
\label{firstpage}
\pagerange{\pageref{firstpage}--\pageref{lastpage}}
\maketitle

\begin{abstract}
We report the first instance of an M dwarf/brown dwarf obliquity measurement for the TOI-2119 system using the Rossiter-McLaughlin effect. TOI-2119~b is a transiting brown dwarf orbiting a young, active early M dwarf ($T_{\rm{eff}}$ = 3553~K). It has a mass of 64.4~M$_{\rm{J}}$ and radius of 1.08~R$_{\rm{J}}$, with an eccentric orbit ($e$ = 0.3) at a period of 7.2~days. For this analysis, we utilise NEID spectroscopic transit observations and ground based simultaneous transit photometry from the Astrophysical Research Consortium (ARC) and the Las Campanas Remote Observatory (LCRO). We fit all available data of TOI-2119~b to refine the brown dwarf parameters and update the ephemeris. The classical Rossiter-McLaughlin technique yields a projected star-planet obliquity of $\lambda=-0.8\pm1.1^\circ$ and a three-dimensional obliquity of $\psi=15.7\pm5.5^\circ$. Additionally, we spatially resolve the stellar surface of TOI-2119 utilising the Reloaded Rossiter-McLaughlin technique to determine the projected star-planet obliquity as $\lambda=1.26 \pm 1.3^{\circ}$. Both of these results agree within $2\sigma$ and confirm the system is aligned, where TOI-2119~b joins an emerging group of aligned brown dwarf obliquities. We also probe stellar surface activity on the surface of TOI-2119 in the form of centre-to-limb variations as well as the potential for differential rotation. Overall, we find tentative evidence for centre-to-limb variations on the star but do not detect evidence of differential rotation. 
\end{abstract}




\section{Introduction}
As a body transits its host star, a portion of the starlight is blocked from the line of sight and a distortion of the disk-integrated radial velocities (RV) is observed, known as the Rossiter-McLaughlin (RM) effect \citep{rossiter1924detection, mclaughlin1924some}. The RM effect is primarily sensitive to the projected obliquity, $\lambda$ (i.e. the sky-projected angle between the stellar spin axis and planetary orbital plane) and has been observed for hundreds of planetary systems \citep[e.g.,][]{triaud2018rossiter,albrecht2022}. 

The Reloaded RM technique \citep[RRM:][]{cegla2016rossiter} isolates the blocked starlight to spatially resolve the stellar spectrum. The isolated starlight is then modelled to account for the stellar rotation behind the planet and used to derive $\lambda$. If the planet occults multiple latitudes, it can be used to determine the stellar inclination, $i_{\star}$ (by disentangling it from the projected stellar rotational velocity, $v_{\rm{eq}}\sin i_\star$, where $v_{\rm{eq}}$ is the stellar equatorial rotational velocity) and we can then combine this with $\lambda$ to determine the true 3D obliquity, $\psi$ \citep[see][]{doyle2023wasp131}. Additionally, we can use the isolated starlight to account for any centre-to-limb convective velocity variations (CLV) on the stellar surface which can impact the RM effect and consequently the derived projected obliquity \citep[see][]{cegla2016modeling, bourrier2017refined, doyle2022WASP-166}.

To date, there are only 46 known transiting brown dwarfs, with $\sim$10 of these being hosted by M dwarfs \citep[see][for examples]{henderson2024ngts, jackman2019ngts, palle2021espresso, acton2021ngts}. This fraction of transiting brown dwarfs is unusually small given there are $\sim$4,300\footnote{\url{https://exoplanetarchive.ipac.caltech.edu/docs/counts_detail.html}{}} known transiting exoplanets. For known M dwarf/brown dwarf systems, there are currently no published obliquity measurements with only five obtained for M dwarf/planet systems. Three of these belong to the TRAPPIST-1 system \citep{brady2023measuring} which are all aligned, and overall amongst the other systems there is a range of obliquities, both aligned and missaligned, but all measurements are for terrestial planets. On the other hand, the number of known M dwarfs hosting hot Jupiters is $\sim$20 \citep[e.g.][]{bayliss2018ngts,gems2024} with no known obliquity measurements in the literature. Therefore, studying the obliquities of M dwarf/brown dwarf systems will help us understand the formation of M dwarf/hot Jupiter systems. For example, sub-stellar companions in aligned orbits may possess a dynamically gentle planet-disc migration history, whereas misaligned sub-stellar companions may have experienced a more violent migration (e.g. Kozai migration). There is evidence that brown dwarfs and planets might form via different formation channels \citep[see][]{bowler2020population}. Therefore, the obliquity measurements from this study can be used as a way to distinguish (i) in-situ formation followed by smooth migration versus (ii) ``high-eccentricity'' migration involving gravitational scattering followed by tidal orbit truncation \citep[see][]{dawson2018origins}.

Previous studies looking into the obliquities of warm Neptunes around low mass M dwarfs \citep[see][]{bourrier2018orbital, stefansson2022warm} have found these systems to range from polar to largely misaligned to aligned. Similarly, for hot Jupiters a range of obliquities have been observed \citep[see][]{rice2022origins}, many with large misalignments or retrograde orbits. Interactions such as planet-planet scattering \citep{beauge2012multiple}, von Zeipel-Lidov-Kozai oscillations \citep{fabrycky2007shrinking}, and secular chaos \citep{wu2011secular} are often used to explain high obliquities which in turn excite orbital eccentricity. Given the eccentric orbit and the large mass of the companion, it is expected TOI-2119~b may be misaligned \citep{albrecht2022}, raising questions on its formation and evolution. The determination of the obliquity of the TOI-2119 system will aid in filling out the parameter space of known obliquities for cool stars, shedding light on the distribution within this population.

TOI-2119~b is a newly discovered transiting brown dwarf which orbits a young and active M dwarf star \citep{canas2022eccentric, carmichael2022toi}. The brown dwarf has a mass of $M_{\rm{BD}}$ = 64.4~M$_{\rm{J}}$, radius of $R_{\rm{BD}}$ = 1.08~R$_{\rm{J}}$, an orbital period of $P_{\rm{orb}}$ = 7.2~d and is in an eccentric orbit of $e$ = 0.337. The host is an early M dwarf ($T_{\rm{eff}}$ = 3553~K) which is bright in the near-infrared and visible with $J$ = 8.976 and $V$ = 12.37. 

In this paper we determine the stellar obliquity for TOI-2119 using NEID spectroscopy observations and simultaneous photometry detailed in \S \ref{sec:obs}. In \S \ref{sec:classical} we detail the classical RM analysis and in \S \ref{sec:reloaded} the reloaded RM analysis. Finally, in \S \ref{sec:con} we discuss the insights into the formation of M dwarf/brown dwarf systems this measurement yields. 

\section{Observations}
\label{sec:obs}
In this study, we use both spectroscopic and photometric data to analyse the TOI-2119 system. In this section, each of the data sources are detailed along with the reduction pipelines and software used. 

\begin{table}
    \centering
    \caption{Summary of the data used in this work.}
    {\sl In-transit Spectroscopic Data: NEID}
    \begin{tabular}{cccccc}
    \hline
    \hline
    Run & Night & $N_{\rm{obs}}$ & $t_{\rm{exp}}$ & $\gamma$$^a$ & SNR$^b$ \\
    & & & (s) & (kms$^{-1}$) & (750~nm) \\
    \hline
    A & 10 May 2023 & 22 & 600 & $-$15.2850 & 25  \\
    B & 15 Jun 2023 & 22 & 600 & $-$15.2848 & 32  \\
    \hline 
    \end{tabular}

    \vspace{5mm}
    {\sl Out-of-transit Spectroscopic Data}
    \begin{tabular}{cccc}
    \hline
    \hline
    Facility & Date & $N_{\rm{obs}}$ & $t_{\rm{exp}}$ \\
    & & & (s) \\
    \hline
    APOGEE-2N & 11 May - 12 May 2018 & 2 & 1342/2684 \\
    HPF & 19 Sep 2020 - 26 May 2021 & 8 & 945 \\
    TRES & 05 Sep - 13 Oct 2020 & 8 & 1200-2800 \\
    NEID & 16 Apr 2022 & 7 & 600 \\
    \hline 
    \end{tabular}
    
    \vspace{5mm}
    {\sl Photometric Data}
    \begin{tabular}{ccccc}
    \hline 
    \hline 
    Facility & Date & $N_{\rm{obs}}$ & $t_{\rm{exp}}$ & $\sigma_{\rm{residual}}$ \\
    & & & (s) & (ppm\,per \\
    & & & & 2\,min) \\
    \hline
    TESS & 16 Apr - 12 May 2020 (S24) & 16020 & 120 & 1.1 \\
    TESS & 14 May - 8 Jun 2020 (S25) & 17174 & 120 & 0.2 \\
    TESS & 22 Apr - 18 May 2022 (S51) & 44091 & 20 & 6.6 \\
    TESS & 18 May - 13 Jun 2022 (S52) & 62232 & 20 & 7.3 \\
    TESS & 03 May - 21 May 2024 (S78) & 7021 & 120 & 0.9 \\
    ARC & 10 May 2023 & 434 & 25 & 1.4  \\
    TMMT & 14 Apr 2021 & 96 & 100 & 697 \\
    TMMT & 15 May 2022 & 289 & 60 & 296 \\
    TMMT & 10 May 2023 & 185 & 30 & 200 \\
    LCRO & 10 May 2023 & 53 & 180 & 379 \\
    \hline
    \end{tabular}
    
        \vspace{2mm}
     \begin{flushleft}
    {\bf Notes:} $^a$ The error on the systemic velocity, $\gamma$, is 3.5~ms$^{-1}$ and 4.01~ms$^{-1}$ for run A and run B respectively. $^b$ The SNR per-pixel was computed as the average SNR for all observations at 750~nm processed by the NEID pipeline.
    \end{flushleft}
    \label{tab:observations}
\end{table}

\begin{figure*}
    \centering
    \includegraphics[width = 0.97\textwidth]{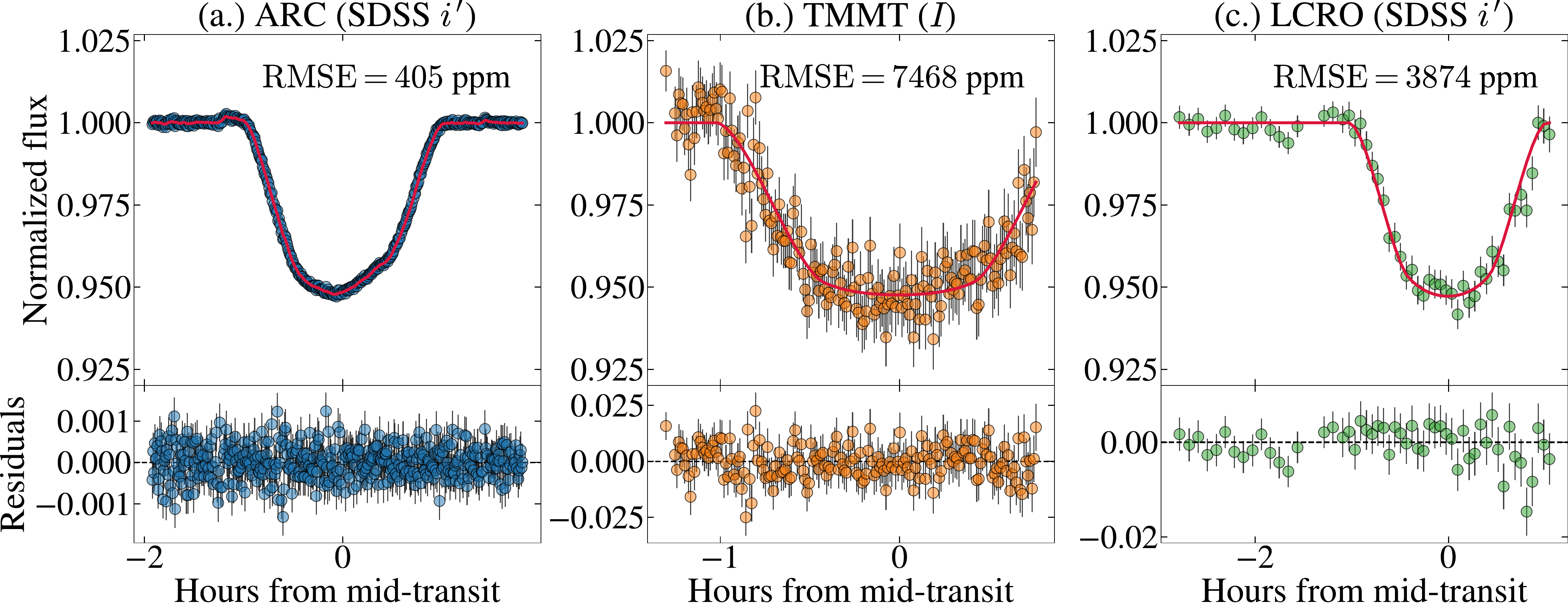}
    \caption{The transit photometry for observations simultaneous to NEID for Run A (10 May 2023) only. {\it Top:} The phase-folded photometry for each instrument along with the best-fit model (solid red line). 
    {\it Bottom:} The residuals to the best-fitting model.}
    \label{fig:phaselc}
\end{figure*}

\subsection{Spectroscopic Data}
We observed two transits (ID: 2023A-547291, PI: L. Doyle) and a pre-transit time series of TOI-2119~b (ID: 2022A-802765, PI: C. Ca\~nas) using the NEID spectrograph on the WIYN\footnote{The WIYN Observatory is a joint facility of the NSF's National Optical-Infrared Astronomy Research Laboratory, Indiana University, the University of Wisconsin-Madison, Pennsylvania State University, Purdue University and Princeton University} 3.5 m Telescope at Kitt Peak Observatory \citep{schwab2016design}, on 10 May 2023 (run A), 15 June 2023 (run B) and 16 April 2022 (pre-transit). NEID is an environmentally stabilized \citep{stefansson2016versatile, robertson2019ultrastable} high-resolution spectrograph (R $\approx$ 110,000). It has an extended red wavelength coverage from 380 -- 930~nm and is a fiber-fed system similar to the Habitable-zone Planet Finder \citep[see][]{kanodia2018overview, kanodia2023stable}. All observations used a fixed exposure time of 600~s. For each transit of TOI-2119~b, we obtained a sequence of 22 observations to reach a signal-to-noise (SNR) per pixel near 27 at 750~nm and achieve a maximum number of in-transit points during the $\sim$2~hr transit duration. Each run respectively covered a duration of 3h 59m and 3h 56m covering the full transit duration and includes $\sim$1~hr pre- and $\sim$1~hr post- baseline. During both runs, there was a period of time during the post- and pre-transit, respectively, where the instrument was performing a routine nightly intermediate calibration sequence. This did not affect the observations during the transit in any way but reduced the number of out-of-transit observations by one on each night. A summary of the NEID observations can be found in Table \ref{tab:observations}. 

The spectra were reduced with version 1.3.2 of the NEID data reduction pipeline\footnote{\href{https://neid.ipac.caltech.edu/docs/NEID-DRP/}{https://neid.ipac.caltech.edu/docs/NEID-DRP/}} (DRP), using an M2 binary mask from ESPRESSO to cross-correlate the observed spectra to generate high SNR cross-correlation functions (CCFs), which we used for our analysis. For the RRM analysis, we utilise a combined CCF for all spectral orders, therefore, we compute this as a weighted sum of the NEID CCF orders. For the classical RM analysis, we used a modified version of the Spectrum Radial Velocity Analyzer \citep[\texttt{SERVAL}:][]{zechmeister2018spectrum} pipeline optimised for NEID spectra \cite[see][for details]{stefansson2022warm}, to extract RVs in the wavelength range from $446-896$~nm (order indices $37-104$). We use the \texttt{SERVAL} RVs for the classical RM analysis because the current DRP masks do not span the entire NEID bandpass. Template-matching algorithms, such as \texttt{SERVAL}, have been shown to provide improved RV precision, particularly for M dwarfs, due to inclusion of the red end of the spectra \citep[e.g.,][]{stefansson2022warm,Canas2022}. For comparison, the median DRP RV precision was 7.2, 10.3, and 7.8 $\mathrm{ms}^{-1}$ compared to the \texttt{SERVAL} RV precision of 3.1, 5.0, and 3.3 $\mathrm{ms}^{-1}$ for the NEID observations on April 2022, May 2023, and June 2023, respectively. The \texttt{SERVAL} calculated NEID RVs are listed in Table \ref{tab:neidrvs}. 

The APOGEE-2N and HPF RVs used in this work are described in detail in \citet{canas2022eccentric} while the TRES RVs are described in \citep{carmichael2022toi}. Conditions for all observations were favourable, with SNR remaining consistent and airmass not exceeding $\sim$1.3. The average integrated radial velocity uncertainties for run A and run B are 4.9~ms$^{-1}$ and 3.5~ms$^{-1}$ respectively. In run A, the SNR can be seen to increase over the duration of the night, which correlates with the decreasing airmass as observing conditions improve. In run B, the SNR increases sharply at the beginning of the night, where it then remains relatively stable throughout the transit observation. Furthermore, the FWHM and contrast remain steady during both runs and are dispersed around the mean. 

\subsection{Photometric Data}
Simultaneous photometry to the NEID spectroscopic transits is vital for determining if there is any contamination of stellar activity (e.g. spots and flares) during these observations \citep[see][]{ciceri2013simultaneous}. Additionally, obtaining an updated ephemeris is important for the RRM analysis in this paper. Therefore, we utilised a total of 14 photometric transits, 11 transits from {\tess}, one transit from ARC, one from TMMT and one from LCRO to determine the properties of the transiting brown dwarf TOI-2119~b. Transits from ARC, TMMT, and LCRO were simultaneous to the 10th May 2023 transits; unfortunately, none were obtained for the 15th June 2023 due to weather issues. Full details of each transit are provided in the following sections. 

\subsubsection{TESS Photometry}
The Transiting Exoplanet Survey Satellite \citep[{\tess}:][]{ricker2014tess} observed TOI-2119 in 2-min cadence capturing a total of 11 transits during Sectors 24, 25, 51, 52, and 78 between the 16 April 2020 and 21 May 2024 (see Table \ref{tab:observations}). We accessed the short cadence light curves produced by the TESS Science Processing Operations Centre (SPOC) pipeline \citep{jenkins2016spoc} and used the \textsc{PDCSAP\_FLUX} time series for our analysis. Observations with non-zero data quality flags \citep[see Table 28 in][for more details]{tenenbaum2018tess} are excluded from further analysis. There are various flares in the light curve of TOI-2119 and, to remove the largest flares, we reject any median normalised observation with a normalised flux larger than 1.01. \citet{canas2022eccentric} and \citet{carmichael2022toi} determined TOI-2119 to have a stellar rotation period of $\sim$13.2~d as well as flaring activity present throughout observations. Therefore, we utilise a Gaussian process in our light curve model (as detailed in \S \ref{sec:classical}).

\subsubsection{Astrophysical Research Consortium  Photometry}
We observed one full transit of TOI-2119~b on 10 May 2023 with the Astrophysical Research Consortium (ARC) 3.5 m Telescope at the Apache Point Observatory (APO) in New Mexico. We used the optical CCD Camera ARCTIC \citep{huehnerhoff2016astrophysical} equipped with an engineered diffuser \citep{stefansson2017toward} which enables near photon/scintillation-limited precision light curves by spreading the stellar point-spread function into a stable top-hat profile without the need to defocus the telescope. The observation had a setup of quad and fast read-out mode, 4 $\times$ 4 pixel binning, 25~s exposures, with biases and dome flats collected before each observing run. ARCTIC does not experience significant dark current for exposures $<$~60~s. For this transit we used the Sloan $i^\prime$ filter and sky conditions were photometric throughout the night.

We reduce each observation with bias subtraction before dividing by a nightly median-combined normalised flat field. Aperture photometry was applied using {\tt AstroImageJ} \citep{astroimagej} assuming an aperture size of 20 pixels (9.1"), five reference stars, and background annuli of 25 (11.4") and 30 (14.7") pixels for inner and outer radii, respectively. We detrend the data by dividing out a linear model calculated from the out-of-transit points. The ARC light curve is shown in Figure \ref{fig:phaselc}. 

\subsubsection{TMMT Photometry}
Three transits of TOI-2119~b were observed using the robotic Three-hundred MilliMeter Telescope \citep[TMMT:][]{monson2017standard} at Las Campanas Observatory (LCO) on the nights of April 13th 2021, May 13th 2022, and May 9th 2023(simultaneous to NEID). All TMMT data processed and the light curve generated following the steps described in the TOI-2119~b discovery paper \citet{canas2022eccentric}, where the April 2021 transit was initially published. The observations were performed slightly out of focus in the Bessell I filter \citep{bessell1990ubvri} with a point spread function FWHM of 4 pixels ($\sim3.9\arcsec$). The observations used an exposure time of 100, 60, and 30~s (plus 13~s readout), respectively, while operating in a 1 $\times$ 1 binning mode. A summary of the TMMT observations can be found in Table \ref{tab:observations} and the TMMT light curve simultaneous to NEID for the night of May 9th 2023 is shown in Figure \ref{fig:phaselc}. 

\subsubsection{Las Campanas Remote Observatory Photometry}
One transit of TOI-2119~b was obtained using the 305 mm Las Campanas Remote Observatory telescope\footnote{\href{http://lcobot.duckdns.org/}{http://lcobot.duckdns.org/}} (LCRO) at LCO on the 10 May 2023. LCRO is an f/8 Maksutov-Cassegrain from Astro-Physics on a German Equatorial AP1600 GTO mount with an FLI Proline 16803 CCD Camera, FLI ATLAS focuser and Centerline filter wheel. The observations were performed slightly out of focus in the Sloan $i'$ filter with a fixed exposure time of 180~s. These observations were carried out using the $1 \times 1$ binning mode, which provide a gain of 1.52 $\mathrm{e^-/ADU}$, a plate scale of \(0.773 \arcsec/\mathrm{pixel}\), and a readout time of 17s. The LCRO transit is shown in Figure \ref{fig:phaselc}.

\begin{table}
    \centering
    \caption{NEID RVs$^a$ used in the classical RM analysis.}
    \begin{tabular}{ccccc}
    \hline
    \hline
    $\mathrm{BJD_{TDB}}$ & RV & $\sigma$ & S/N$^a$ & Run \\
    & ($\mathrm{m~s^{-1}}$) & ($\mathrm{m~s^{-1}}$) & &\\
    \hline
    2459685.883660 &   201.8 &  3.0 & 35.9 & pre-transit \\
    2459685.890883 &   121.5 &  2.8 & 37.9 & pre-transit \\
    2459685.897944 &    33.6 &  3.1 & 35.3 & pre-transit \\
    2459685.905517 &   -46.2 &  2.8 & 38.4 & pre-transit \\
    2459685.912083 &  -125.7 &  3.4 & 32.1 & pre-transit \\
    2459685.919514 &  -214.8 &  5.8 & 21.1 & pre-transit \\
    2459685.926992 &  -284.7 &  4.1 & 27.6 & pre-transit \\
    \hline
    2460074.731809 &   923.2 &  5.0 & 24.2 & A \\
    2460074.739094 &   823.2 &  4.6 & 25.4 & A \\
    2460074.746565 &   737.6 &  4.9 & 24.0 & A \\
    2460074.753950 &   667.6 &  5.5 & 22.1 & A \\
    2460074.761123 &   563.7 &  5.9 & 21.3 & A \\
    2460074.768724 &   500.3 &  5.6 & 21.8 & A \\
    2460074.775911 &   420.4 &  5.9 & 21.1 & A \\
    2460074.783382 &   365.6 &  6.5 & 19.5 & A \\
    2460074.790834 &   289.7 &  5.4 & 22.6 & A \\
    2460074.798064 &   199.2 &  4.6 & 25.4 & A \\
    2460074.805648 &   101.6 &  4.3 & 27.2 & A \\
    2460074.813031 &     1.6 &  4.0 & 28.5 & A \\
    2460074.820371 &   -97.1 &  3.7 & 29.8 & A \\
    2460074.827773 &  -189.8 &  3.8 & 29.1 & A \\
    2460074.835047 &  -274.3 &  3.7 & 29.5 & A \\
    2460074.842375 &  -332.2 &  4.1 & 27.5 & A \\
    2460074.849926 &  -398.7 &  4.5 & 25.7 & A \\
    2460074.857087 &  -466.1 &  4.8 & 23.9 & A \\
    2460074.864361 &  -549.9 &  5.6 & 21.6 & A \\
    2460074.882764 &  -736.4 &  5.1 & 22.7 & A \\
    2460074.889894 &  -808.8 &  5.3 & 22.2 & A \\
    2460074.897427 &  -880.4 &  5.1 & 23.0 & A \\
    \hline
    2460110.734735 &  1052.0 &  4.1 & 26.3 & B \\
    2460110.742332 &   964.9 &  4.5 & 24.7 & B \\
    2460110.749558 &   885.5 &  4.0 & 27.1 & B \\
    2460110.770476 &   649.8 &  5.4 & 21.2 & B \\
    2460110.777771 &   576.1 &  3.9 & 27.7 & B \\
    2460110.785493 &   512.0 &  3.8 & 28.6 & B \\
    2460110.792606 &   442.4 &  3.4 & 31.0 & B \\
    2460110.799714 &   359.5 &  3.5 & 30.0 & B \\
    2460110.807270 &   265.0 &  3.2 & 32.6 & B \\
    2460110.814381 &   160.3 &  3.1 & 33.6 & B \\
    2460110.820684 &    71.8 &  3.2 & 33.1 & B \\
    2460110.827399 &   -10.8 &  3.0 & 34.6 & B \\
    2460110.834545 &   -97.5 &  3.3 & 31.9 & B \\
    2460110.840987 &  -164.7 &  3.2 & 32.2 & B \\
    2460110.847689 &  -219.2 &  3.0 & 34.0 & B \\
    2460110.854873 &  -283.3 &  3.2 & 32.8 & B \\
    2460110.862335 &  -356.1 &  3.3 & 31.9 & B \\
    2460110.869373 &  -430.1 &  3.3 & 32.6 & B \\
    2460110.876625 &  -508.4 &  3.2 & 33.0 & B \\
    2460110.884123 &  -589.3 &  3.4 & 31.3 & B \\
    2460110.891343 &  -661.2 &  3.6 & 30.1 & B \\
    2460110.898821 &  -735.3 &  3.3 & 31.9 & B \\
    \hline 
    \end{tabular}
    \vspace{2mm}
    \begin{flushleft}
    {\bf Notes:} $^a$ \texttt{SERVAL} derived RVs. $^b$ The S/N is the median value per 1D extracted pixel at 850 nm for NEID. All exposure times are 600s.
    \end{flushleft}
    \label{tab:neidrvs}
\end{table}

\section{Classical Rossiter McLaughlin}
\label{sec:classical}

We jointly model all photometry (TESS, ARC, TMMT, LCRO) and RV data (NEID RVs are SERVAL derived in the bandpass $446-896$~nm) described in \S \ref{sec:obs} and Table \ref{tab:observations} using a modified version of \texttt{juliet} \citep{juliet2019}. Similar to \citet{canas2022eccentric}, the photometric model accounts for both the transit and occultation of TOI-2119~b. The photometric model was calculated using the \texttt{batman} package \citep{Kreidberg2015}. The transit were calculated with a quadratic limb-darkening law where the limb-darkening coefficients were sampled following \cite{Kipping2013}, not included in Table \ref{planetary_properties}. The occultation model was calculated using a uniform limb darkening law. We accounted for photometric variability in the TESS photometry using a quasi-periodic Gaussian Process noise model that was calculated with the \texttt{celerite} package \citep{Foreman-Mackey2017}. 

\begin{figure}
    \centering
    \includegraphics[width = 0.47\textwidth]{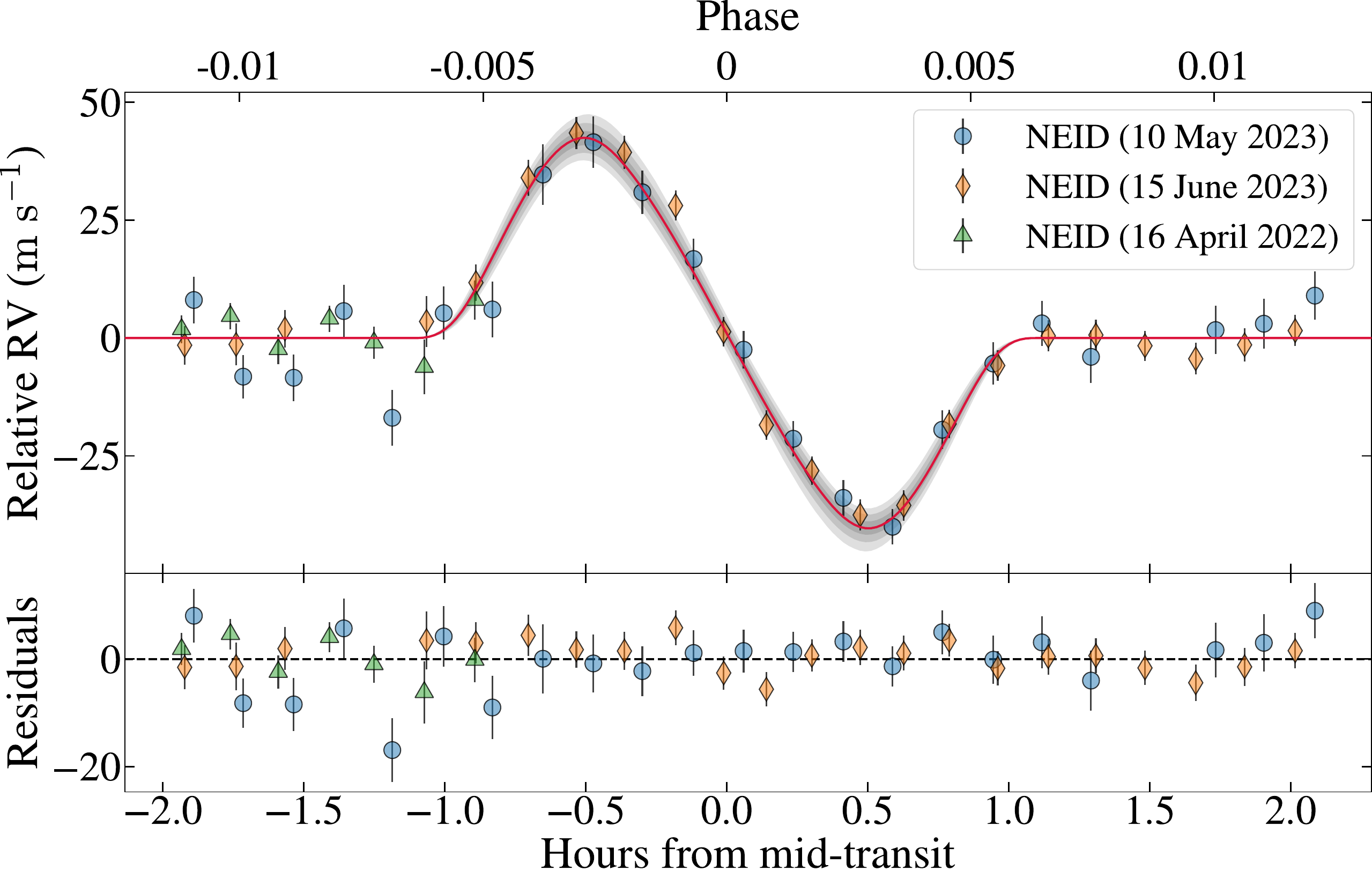}
    \caption{The classical RM effect for TOI-2119~b. {\it Top:} The phase-folded NEID RVs, after subtracting the RV offsets and Keplerian orbit, along with the best-fit model (solid red line) and the $1-3\sigma$ range of our models (gray shaded regions). 
    {\it Bottom:} The residuals to the best-fitting model.}
    \label{fig:classicalrm}
\end{figure}

The RV model is the sum of (i) a standard Keplerian model calculated using the \texttt{radvel} package \citep{Fulton2018} and (ii) the RM effect model for a Gaussian spectral line profile from \citet{Hirano2011} that was calculated using the \texttt{rmfit}\footnote{\href{https://github.com/gummiks/rmfit}{https://github.com/gummiks/rmfit}} package \citep{rmfit2020}. We follow the methodology of \citet{stefansson2022warm} to constrain $\psi$ by adopting normal priors on the known rotation period ($P_\mathrm{rot}=13.2\pm0.2$ days) and stellar radius ($R_\star=0.51\pm0.01~\mathrm{R_\odot}$) from \citet{canas2022eccentric} along with placing uniform priors on the stellar inclination ($\cos i_\star$) and projected stellar obliquity ($\lambda$). The RM effect model used a quadratic limb-darkening law and, like the transit model, included a uniform prior on the limb-darkening coefficients following \citet{Kipping2013}. Similar to \citet{stefansson2022warm}, we set a prior on the Gaussian dispersion of the spectral line in the RM effect model based on the width of the NEID resolution element ($\beta=3.0\pm0.5~\mathrm{km~s^{-1}}$). We accounted for the finite duration of our observations by super-sampling $N=30$ data points at each time step and averaging those to generate the model. Each RV time series included a different offset. All the photometric and RV models also included a simple white-noise model in the form of a jitter term that was added in quadrature to the error bars of each instrument. We performed the parameter estimation using \texttt{dynesty} \citep{Speagle2020}. The best-fit and the model posteriors are displayed in Figure \ref{fig:classicalrm}. The modelling reveal that TOI-2119 is well-aligned with a projected stellar obliquity of $\lambda=-0.8\pm1.1~^\circ$ and a three-dimensional obliquity of $\psi=15.7_{-5.6}^{+5.4}~^\circ$.

\begin{table}
    \centering
    \caption{System parameters for TOI-2119}
    \begin{tabular}{l|c}
    \hline 
    Parameter (unit) & Value \\
    \hline 
    \multicolumn{2}{l}{\bf Stellar parameters from \citet{canas2022eccentric}}\\
    \hline
    $T_{\rm eff}$ (K) & $3553 \pm 67$\\
    ${\rm [Fe/H]}$ & $0.1 \pm 0.1$\\
    $\log g_\star$ & $4.74 \pm 0.04$\\ 
    $M_\star$ ($M_{\rm \odot}$) & $0.53 \pm 0.02$\\
    $R_\star$ ($R_{\rm \odot}$) & $0.51 \pm 0.01$\\
    Age (Gyr) & 0.7 – 5.1 \\
        \hline 
    \multicolumn{2}{l}{\bf Planetary parameters from \citet{canas2022eccentric}} \\
    \hline
    $T_{\rm{0}}$ (BJD) & $2458958.67756 \pm 0.000064$ \\
    $P$ (days) & $7.200861 \pm 0.000005$ \\
    $T_{\rm{dur}}$ (hours) & $2.039 \pm 0.007$ \\
    $R_{\rm{BD}} / R_{\star}$ & $ 0.226 \pm 0.001$ \\
    $b$ & $0.623 \pm 0.009$ \\
    $e$ & $0.3362 \pm 0.0005$ \\
    $\omega$ ($^{\circ}$) & $-0.9 \pm 0.03$ \\
    $K$ ($\mathrm{km~s}^{-1}$) & $10.59 \pm 0.02$\\
    $a/R_\star$ & 27.0 $\pm$ 0.2 \\
    $i_{\rm{BD}}$ ($^{\circ}$) & $88.51 \pm 0.03$ \\
    \hline
    \hline 
    \multicolumn{2}{l}{\bf Fitted Planetary parameters from this work} \\
    \hline
    $T_{\rm{0}}$ (BJD) & $2458958.67837 \pm 0.00004$ \\
    $P$ (days) & $7.2008569 \pm 0.0000003$ \\
    $T_{\rm{dur}}$ (hours) & 2.040 $\pm$ 0.005 \\
    $R_{\rm{BD}} / R_{\star}$ & $0.2221 \pm 0.0007$ \\
    $b$ & $0.639 \pm 0.005$ \\
    $e$ & $0.3355 \pm 0.0002$ \\
    $\omega$ ($^{\circ}$) & $-0.94 \pm 0.09$ \\
    $K$ ($\mathrm{m~s}^{-1}$) & $10584.1_{-8.3}^{+7.7}$\\
    $a/R_\star$ & 26.7 $\pm$ 0.1 \\
    $i_{\rm{BD}}$ ($^{\circ}$) & $88.47 \pm 0.02$ \\
    \hline
    {\it Classical RM Parameters:} & \\
    \hline
    $u_{1,\mathrm{NEID}}$ & $0.77_{-0.28}^{+0.27}$ \\
    $u_{2,\mathrm{NEID}}$ & $-0.12_{-0.26}^{+0.31}$ \\
    $\beta$ ($\mathrm{km~s}^{-1}$) & $2.7\pm0.3$ \\
    $v_{\mathrm{eq}}\sin i_\star$ ($\mathrm{km~s}^{-1}$) & $1.92\pm0.06$ \\
    $\cos i_\star$ & $0.29_{-0.10}^{+0.09}$ \\
    $i_\star$ ($^{\circ}$) & $72.9_{-5.4}^{+5.7}$ \\
    $\lambda$ ($^{\circ}$) & $-0.8\pm1.1$ \\
    $\psi$ ($^{\circ}$) & $15.7_{-5.6}^{+5.4}$ \\
    \hline
    {\it Limb-darkening coefficients$^a$ (RRM analysis):} & \\
    \hline
    $u_1$ & \Nuonets \\
    $u_2$ & \Nutwots \\
    \hline
    \end{tabular}
    \begin{flushleft}
    {\bf Notes:} $^a$ The limb darkening coefficients $u_1$ and $u_2$ were obtained in the NEID passband (380 -- 930~nm) by inputting the TOI-2119 stellar parameters into the ExoCTK  calculator \citep{matthew_bourque}.
    \end{flushleft}
    \label{planetary_properties}
\end{table}

\section{Reloaded Rossiter McLaughlin}
\label{sec:reloaded}
We utilised the Reloaded RM (RRM) technique to isolate the starlight of TOI-2119 behind the brown dwarf during its transit. A comprehensive description of the technique can be found in \citet{cegla2016rossiter, doyle2022WASP-166, doyle2023wasp131} and we will include a short description here. In this section, we will use the term \locccf to refer to the occulted light emitted behind the brown dwarf and the term \DIccf to refer to the light emitted by the entire stellar disk. 

\begin{figure}
    \centering
    \includegraphics[width = 0.47\textwidth]{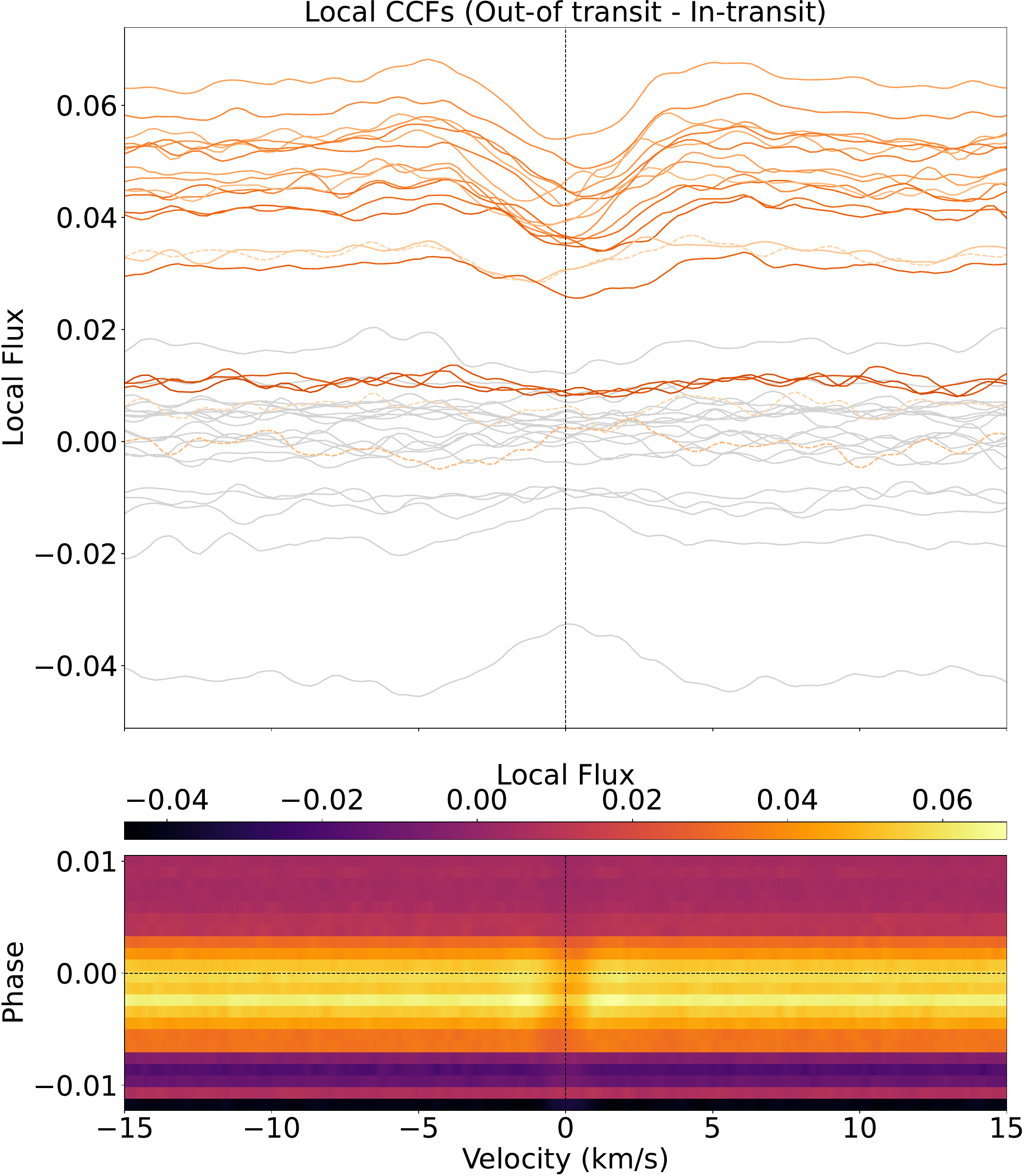}
    \caption{{\it Top:} The local CCFs (out-of-transit – in-transit CCFs) of the star behind the transiting brown dwarf TOI-2119~b in the stellar rest frame. The light grey are the out-of-transit observations and the orange lines are the in-transit observations. The changing gradient of the orange lines represents the changing centroid position where the darker orange is more redshifted. Dashed lines are observations which have a stellar disk position $\langle\mu\rangle <$ 0.40 and are not used in the analysis. A dotted line at 0~kms$^{-1}$ is included to guide the eye. 
    {\it Bottom:}  A top view of the top plot showing a map of the local CCFs colour-coded by the local flux. A dotted line at phase zero and 0~km s$^{-1}$ in both plots is included to guide the eye. }
    \label{local_ccfs}
\end{figure}

NEID \DIccfs are shifted and re-binned in velocity space to correct for Keplerian motions induced by TOI-2119~b. Master-out \DIccf are created for each run by summing all out-of-transit \DIccfs and normalising the continuum to unity. Fitting a Gaussian profile then determines the systemic velocity ($\gamma$; see Table \ref{tab:observations}) which are subtracted from all \DIccfs to the stellar rest frame. Each \DIccf was normalised by their individual continuum value and scaled using a quadratic limb darkened transit model from the fitted parameters in Table \ref{planetary_properties}. Finally, the \locccfs were obtained by subtracting the now scaled in-transit \DIccfs from the master-out \DIccf for each night, see Figure \ref{local_ccfs}. 

\begin{figure}
    \centering
    \includegraphics[width = 0.47\textwidth]{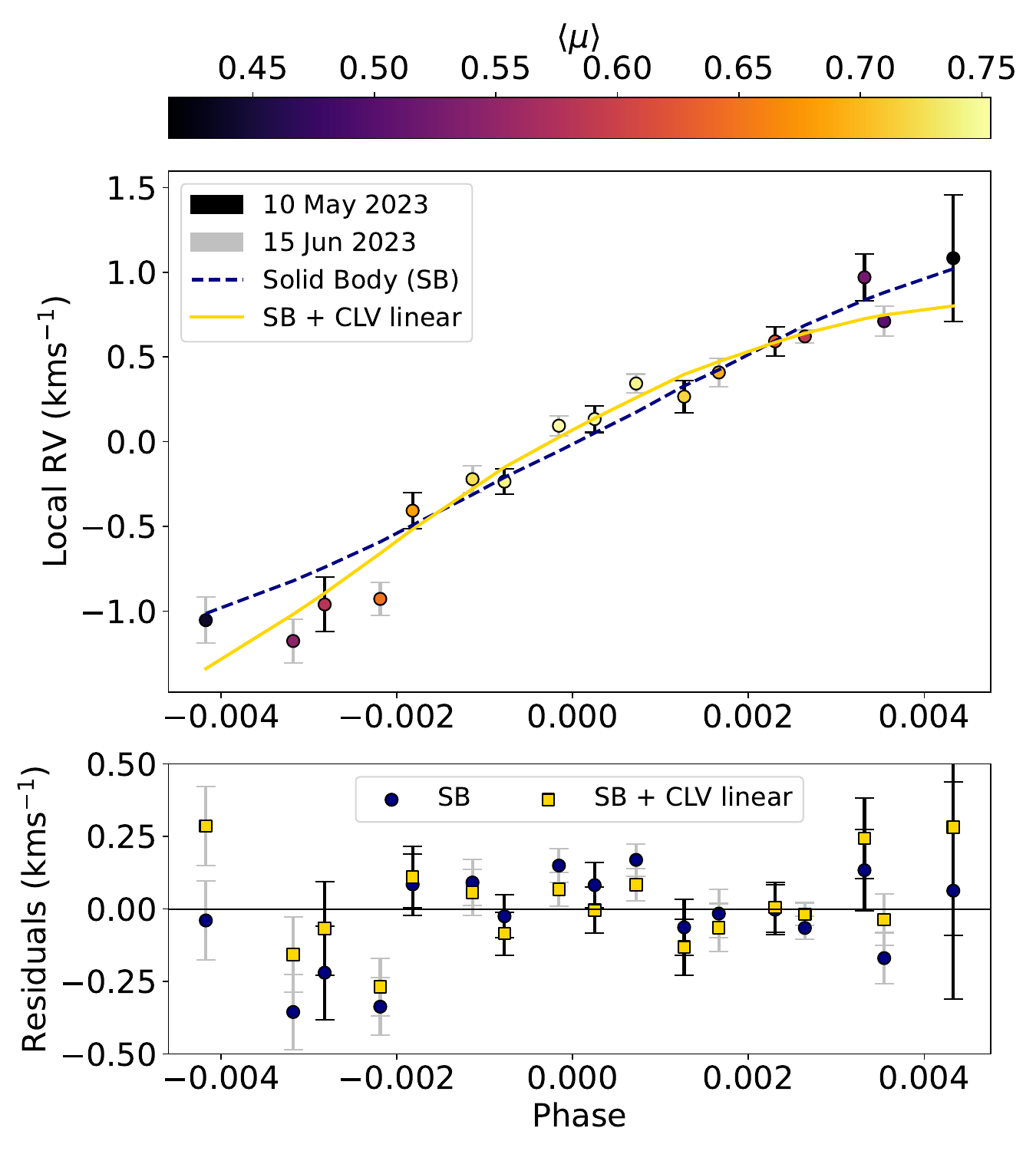}
    \caption{The top panel shows the local RVs determined from the local CCFs of the regions occulted by the brown dwarf as a function of phase. The data points are colour coded by the stellar disk position behind the brown dwarf in units of brightness weighted $\langle\mu\rangle$ (where $\mu = \cos\theta$). The model for Solid Body (SB: navy dashed line) is shown, along with SB plus centre-to-limb linear (gold solid line) model. The bottom panel shows the residuals (local RVs - model) for all models with colours corresponding to the top panel model lines, with a horizontal line at 0 to guide the eye.}
    \label{fig:local_RVs}
\end{figure}

To determine the stellar velocity of the occulted starlight, Gaussian profiles are fitted to each of the \locccfs with parameters including the offset (i.e. continuum), amplitude, centroid, and FWHM. Flux errors assigned to each \locccf were determined as the standard deviation of the CCF continuum and included in our Gaussian fit. Figure \ref{fig:local_RVs} shows the resulting local RVs of the brown dwarf occulted starlight, plotted as a function of both phase and stellar disk position behind the brown dwarf. A total of 4 CCFs with limb angle $\mu < 0.40$ (i.e. the distance from the centre to the limb of the star where $\mu \equiv \cos\theta$ and $\theta$ is the centre-to-limb angle) were removed from our analysis. Profiles close to the limb are very noisy and when comparing the depth to the noise the signal was not significant enough to enable an accurate Gaussian fit, see Figure \ref{local_ccfs} where they are shown as dashed lines.

The local RVs in Figure \ref{fig:local_RVs} were fitted using the model and coordinate system described in \citet{cegla2016rossiter}. This fitting depends on brightness weighted RVs occulted by the brown dwarf, projected obliquity ($\lambda$), stellar inclination ($i_\star$), the equatorial rotational velocity ($v_{eq}$), the differential rotational shear ($\alpha$, ratio between the equatorial and polar stellar rotational velocities), quadratic stellar limb darkening (u$_1$ and u$_2$), and centre-to-limb convective variations of the star ($v_{\rm{conv}}$). For TOI-2119, we fit different scenarios depending on whether or not we account for differential rotation (DR) and centre-to-limb convective variations (CLV) were the results of each are listed in Table \ref{tab:rrm_results}.

\subsection{Results}
The measured local RVs of TOI-2119 increase with orbital phase as the brown dwarf tracks across the stellar disk from approximately $-$1.0~kms$^{-1}$ to +1.0~kms$^{-1}$ (see Figure \ref{fig:local_RVs}), suggesting the TOI-2119 system is likely aligned. Additionally, there appears to be slight curvature within the velocities which suggests the presence of either differential rotation or centre-to-limb convective velocities. Therefore, in this section we discuss the various stellar rotation scenarios we fitted to the local RVs along with exploring the detection of centre-to-limb convective variations.

\begin{table*}
    \centering
        \caption{MCMC observational results for the TOI-2119 system including the derived 3D spin-orbit obliquity.}
    \resizebox{1.0\textwidth}{!}{    
    \begin{tabular}{lccccccccccccc}
    \hline
        Model & No. of Model & $v_{\rm{eq}}$ & $i_{\rm{*}}$ & $\alpha$ & $\lambda$    &  $c_1$        & $c_2$        & $c_3$        & $\Delta$BIC & \chisq &  $\chi_{\nu}^{2}$ &  $\psi$  \\
              & Parameters   & (km s$^{-1}$)  & ($^{\circ}$) &          & ($^{\circ}$) & (km s$^{-1}$) & (km s$^{-1}$)&   (km s$^{-1}$)&     &        &                   &($^{\circ}$) \\
        \hline
        SB & 2 &  $1.51 \pm 0.1$ & 90.0 & 0.0 & $-0.69\substack{+1.3 \\ -1.2}$ & -- & -- & -- & 18.5 & 47.7 & 2.8 &-- \\
        {\bf SB + CLV1} & {\bf 3} & {\bf 1.61 $\pm$ 0.1} & {\bf 90.0} & {\bf 0.0} & {\bf 1.26 $\pm$ 1.3} & {\bf 1.09 $\pm$ 0.2} & {\bf --} & {\bf --} & {\bf 0.0} & {\bf 26.4} & {\bf 1.5} & {\bf --} \\
        SB + CLV2 & 4 & $1.78 \pm 0.1$ & 90.0 & 0.0 & $20.4\substack{+5.8 \\ -6.4}$ & $-9.6 \pm 3.7$ & $-8.4 \pm 2.9$ & -- & -5.4 & 18.2 & 1.07 & -- \\
        SB + CLV3 & 5 &  $1.77 \pm 0.1$ & 90.0 & 0.0 & $16.1\substack{+11.4 \\ -12.5}$ & $-25.2\substack{+37.4 \\ 37.2}$ & $34.5\substack{+62.0 \\ -62.3}$ &$-14.3\substack{+34.1 \\ -34.0}$ & -2.2 & 18.4 & 1.08 & -- \\
        DR &  4 & $2.49\substack{+3.1 \\ -0.8}$ & $61.4 \substack{+81.7 \\ -43.4}$  & $-0.11 \pm 0.6$ & $-0.67 \pm 0.6$ & -- & -- & -- & 182 & 206 & 12.1 &  $27.1\substack{+27.5 \\ -43.0}$ \\
        DR (towards) &  4 & $2.60\substack{+3.4 \\ -0.8}$ & $35.3 \substack{+36.8 \\ -21.3}$  & $-0.06\substack{+0.7 \\ -0.6}$ & $-0.68\substack{+1.3 \\ -1.2}$ & -- & -- & -- & 24.2 & 47.7 & 2.8&  $53.2\substack{+21.3\\ -36.8}$ \\
        DR (away) &  4 & $2.53\substack{+3.9 \\ -1.0}$ & $145.1 \substack{+22.7 \\ -42.0}$  & $-0.17 \pm 0.5$ & $-0.56\substack{+1.4 \\ 1.3}$ & -- & -- & -- & 27.5 & 51.1 & 3.0 &  $56.6\substack{+22.7\\ -41.9}$ \\
        DR + CLV1 & 5 & $2.59 \substack{+3.3 \\ -0.9}$ & $62.9\substack{+81.2 \\ -43.6}$ & $-0.21\substack{+0.6\\ -0.5}$ & $1.25\pm 1.3$ & $1.09 \pm 0.2$ & -- & -- & 193 & 214 & 12.6 & $ 25.6\substack{+30.1 \\ -43.7} $ \\
        DR + CLV1 (towards) & 5 & $2.93 \substack{+3.5 \\ -1.1}$ & $33.2\substack{+34.4 \\ -19.3}$ & $-0.15\substack{+0.7\\ -0.6}$ & $1.25\substack{+1.3 \\ -1.2}$ & $1.08 \pm 0.2$ & -- & -- & 5.6 & 26.3 &  1.5 & $55.3\substack{+19.3 \\ -34.3}$ \\
        DR + CLV1 (away) & 5 & $2.45 \substack{+3.6 \\ -1.0}$ & $144.3\substack{+22.1 \\ -37.3}$ & $-0.29 \pm 0.4$ & $1.23\substack{+1.3 \\ -1.2}$ & $1.08 \pm 0.2$ & -- & -- & 8.3 & 29.0 &  1.7 & $55.8\substack{+22.1 \\ -37.3}$ \\\\
        \hline
    \end{tabular}}
    \label{tab:rrm_results}
    \vspace{2mm}
     \begin{flushleft}
   {\bf Notes:} For all SB models $i_\star$ and $\alpha$ are fixed under the assumption of rigid body rotation and the $v_{\rm{eq}}$ column corresponds to $v_{\rm{eq}}\sin i_\star$. For these models we are unable to determine the 3D obliquity, $\psi$. The BIC of each model was calculated using \chisq and the reduced chi-squared ($\chi_{\nu}^{2}$) has been added as well for completeness. In each instance we show the $\Delta$BIC where the best fit model is set to zero. For clarity, CLV1, CLV2 and CLV3 correspond to centre-to-limb linear, quadratic and cubic respectively. The best fit model has been highlighted in bold. Corner plots for MCMC runs are available as supplementary material online.    
    \end{flushleft}
    \label{mcmc_fits}
\end{table*}

\subsubsection{Stellar Solid Body Rotation}
\label{sec:sb}
We fit a solid body (SB) stellar rotation model, the simplest of models with the least free parameters, $\lambda$, and $v_{\rm{eq}}\sin i_\star$. Further to this, we are interested in how the net convective blueshift varies across the stellar disk, see \S \ref{sec:clv} for more details. Therefore, to model these centre-to-limb convective variations (CLV) we fit the local RVs for both SB and CLV at the same time. Since we do not know the shape of the trend of the CLV, we firstly test a linear, quadratic or cubic polynomial as a function of limb angle. 

The results for the CLV model fits can be found in Table \ref{tab:rrm_results}, along with the Bayesian Information Criterion (BIC) for each of the models. Overall, a difference in BIC of $\sim$6 between models signifies strong evidence of the lower BIC model being the better fit to the data \citep{raftery1995bayesian,lorah2019value}. Therefore, fitting both NEID transits together and according to the BICs between the SB models, we conclude the best fit to the data is the SB plus a linear CLV. The SB plus quadratic and cubic models have BIC values which are lower than the SB plus linear model but they are not at the threshold for being significant given the additional free parameters. Additionally, there are strong correlations present between the projected obliquity and the centre-to-limb coefficients in the SB plus CLV quadratic and cubic models.  Therefore, with this in mind we select the SB plus linear CLV model as the preferred fit to, giving $v_{\rm{eq}}\sin i_\star = 1.61 \pm 0.1$~km s$^{-1}$ and $\lambda = 1.26 \pm 1.3^{\circ}$ and $c_1 = 1.09 \pm 0.2$~km s$^{-1}$. This is the first measurement of $v_{\rm{eq}}\sin i_\star$ and the projected obliquity for this system, which shows an alignment. The classical RM computed in this work does not take into account differential rotation or centre-to-limb effects, so, when comparing only the SB RRM case the obliquities agree to within 2$\sigma$. 

\subsubsection{Stellar Differential Rotation}
It is well known that the Sun rotates differentially, with the angular velocity decreasing from the equator to the poles. For the Sun, the horizontal differential shear, $\alpha$ = 0.2 and the shear between the pole and the equator, $\Delta\Omega$ = 0.07~rad~d$^{-1}$ with rotation periods of 35~d and 25~d respectively \citep{thompson1996differential}. Stellar differential rotation (DR) has also been observed in early M dwarf stars with a differential rotational shear of $\alpha$ between 0.06 and 0.12 \citep{donati2008large}. In a more detailed study, \citet{zaleski2020activity} detected DR in Kepler-45 using starspot observations in {\sl Kepler} light curves, finding a horizontal shear of $\alpha$ = 0.07 and $\Delta\Omega$ = 0.03~rad~d$^{-1}$. Similarly, \citet{donati2008large} complete a spectroscopic survey of M dwarfs using the Zeeman effect to detect DR in several M dwarfs including CE Bo which has an $\alpha$ and $\Delta\Omega$ comparable to that of Kepler-45. 

If differential rotation is present and TOI-2119~b occults multiple stellar latitudes, we can determine the true 3D obliquity through disentangling $v_{\rm{eq}}\sin i_\star$. Therefore, in addition to solid body rotation, we fit a model to the local RVs assuming DR, where the model parameters are $\alpha$, $\lambda$, $v_{\rm{eq}}$ and $i_\star$. We assume a differential rotation law derived from the Sun following Equation 8 of \citet{cegla2016rossiter}, where we refer the reader also for full details on the coordinate system and equations used. Based on the results for the SB models, we also accounted for centre-to-limb convective variations and fit the local RVs for both DR a linear CLV contribution at the same time. This is important because the rotational shear could be on the same order of magnitude as the limb-dependent convective variations.

Unfortunately, we have bimodal distributions present in both $\alpha$ and $i_\star$, informing us that there remains a degeneracy within them. This could be due to the spectroscopic transits not being precise enough to distinguish whether the star is pointing away from or towards us. Therefore, we ran each DR and DR plus linear CLV model (since this was the best fit amongst the SB models) fixing $i_\star$ < 90$^{\circ}$ (towards) and $i_\star$ > 90$^{\circ}$ (away) to get an estimate on $i_\star$ and $\alpha$. \citet{canas2022eccentric} estimate a stellar rotation period of 13.2~d from rotational modulation present in {\it TESS} light curves. From this we can disentangle $i_\star$ from $v_{\rm{eq}}\sin i_\star$ by using the rotation period and our $v_{\rm{eq}}\sin i_\star =$ 1.61~kms$^{-1}$, yielding $i_\star\sim$55.4$^{\circ}$. We can then use this to predict the stellar latitudes transited by the brown dwarf, using P$_{\rm{rot}}$ and $v_{\rm{eq}}\sin i_\star$, which gives a $\sim$10$^{\circ}$ change in latitude. 

Overall, out of all the DR models, the DR plus linear CLV for the towards case has the lowest BIC and is the preferred model. This yields an $i_\star$ = 33.2$^{\circ}$ which is within the errors when compared to the disentangled $i_\star$ using the stellar rotation period. However, this BIC is still approximately greater than six when compared to the SB plus linear CLV case, so the SB plus CLV linear model is preferred overall. 

\begin{figure}
    \centering
    \includegraphics[width = 0.47\textwidth]{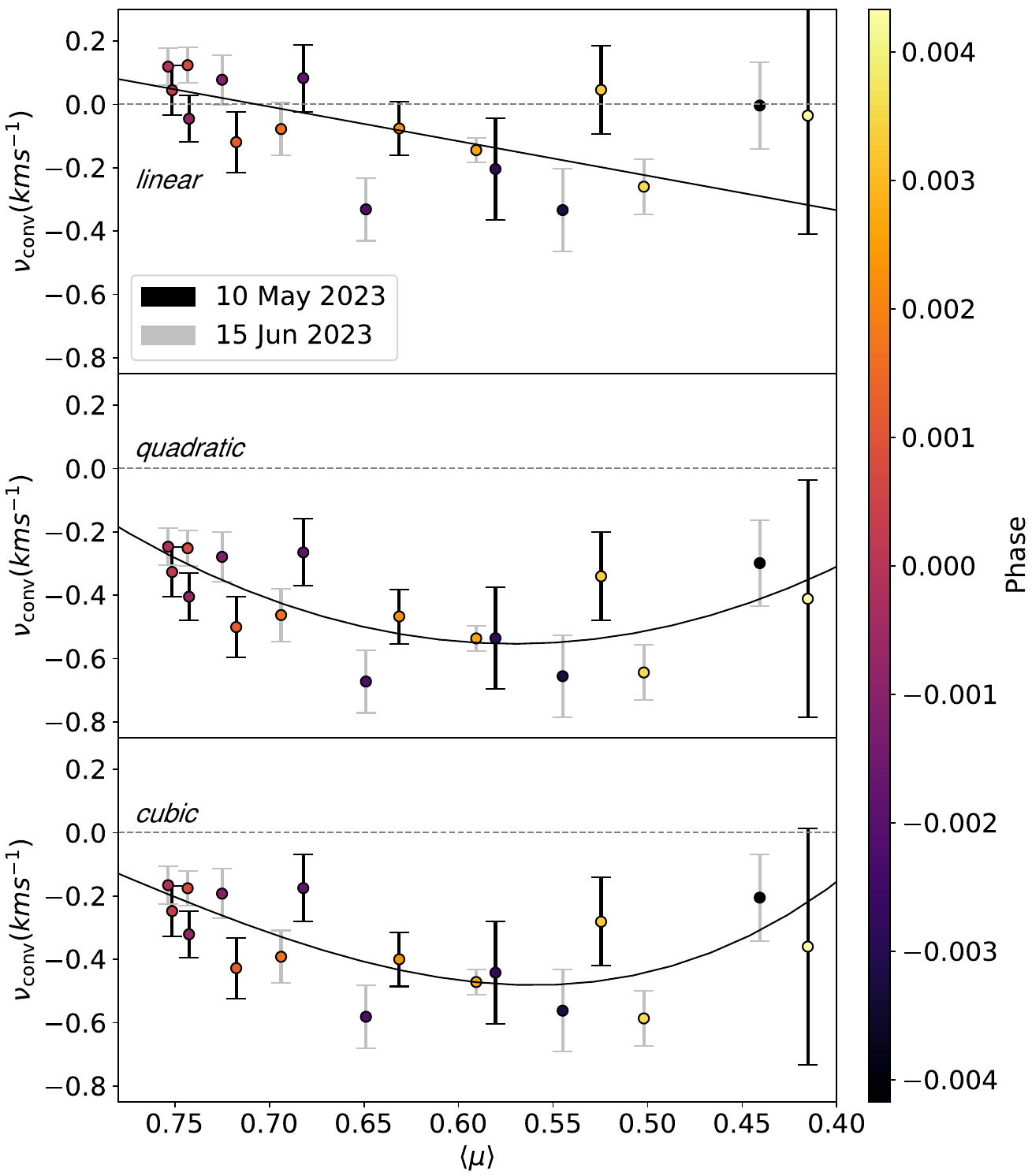}
    \caption{The net convective shifts determined by subtracting the solid body model fit (which changes slightly when adding in CLV) from the local RVs of the in-transit local CCFs, plotted as a function of stellar disk position behind the brown dwarf (brightness weighted $\langle\mu\rangle$). Model fits to the velocities are plotted as solid body linear (top), solid body quadratic (middle) and solid body cubic (bottom). Horizontal grey dashed lines are plotted at $v_{\rm{conv}}$ = 0 to guide the eye and the points are colour coded according to phase.}
    \label{fig:clv}
\end{figure}

\subsubsection{Centre-to-limb Convective Variations}
\label{sec:clv}
As mentioned in \S \ref{sec:sb}, we are interested in how the net convective blueshift varies across the stellar disk. The net convective velocity shift caused by granules changes as a function of limb angle (i.e. from the centre to the limb of the star) due to line-of-sight (LOS) changes. This centre-to-limb convective blueshift has been observed on the Sun with RVs changing on the level of 100~ms$^{-1}$ \citep[see][]{dravins1982photospheric} and on other stars including the K-dwarf HD~189733 \citep[see][]{czesla2015center, cegla2016rossiter}. Therefore, for each of the SB and DR stellar rotation scenarios, we also fit for centre-to-limb convective variations (CLV) including a linear, quadratic and cubic trend, see Table \ref{tab:rrm_results}. We find the most preferred model is the SB plus linear CLV and show the corresponding CLV fits in Figure \ref{fig:clv}. It is clear there is a linear trend present in the velocities once the contribution from SB rotation has been removed. This also manifests as a curvature in the local RVs which we see in Figure \ref{fig:local_RVs}. 

However, the presence of CLV in M dwarfs is still largely debated and currently has not been resolved. In \cite{beeck2013three} they use 3D hydrodynamic simulations to model convection in a range of cool main sequence stars. They then use these models to synthesise Fe~I line profiles, finding all stars except M2 spectral type show a pronounced increasing redshift at decreasing $\mu$. They conclude the for stars with decreasing effective temperature, the transition from convective to radiative heat transport occurs over a greater range of pressure scale heights
and sets in at greater optical depth. This causes an optically thick convectively stable layer which veils granules, breaking convective flows and smearing out inhomogeneties as the lanes and granules do not have as much contrast. In another study, \cite{liebing2021convective} develop a new method for measuring convective blueshift through a model of line core shift as a function of line depth. They measure the strength of convective blueshift for 810 stars observed by HARPS spanning spectral types late-F, G, K and early-M. Overall, they find the strength of convective blueshift scales strongly with effective temperature above 4,700~K. For stars with effective temperature between 4,100~K and 3,800~K, convective blueshift shows a sharp decline and remains constant around zero. Therefore, this in combination with the findings from \citet{beeck2013three} suggest there will not be a strong presence of CLV effects in M dwarfs. 

\begin{figure}
    \centering
    \includegraphics[width=0.47\textwidth]{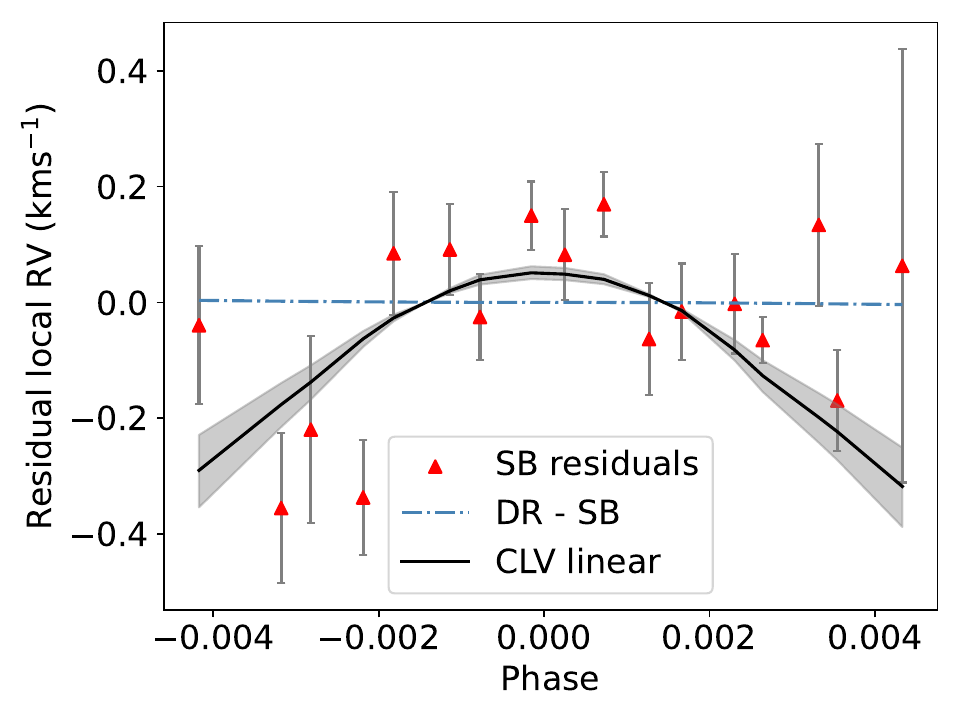}
    \caption{The solid black line represents the linear CLV contribution from the SB plus linear CLV model fit, where the grey shaded region represents the errors. The dashed blue line is the SB model subtracted from the DR model, the errors are large and extend out of the plotted window. Finally, the residuals of the SB model are plotted as red triangles with their corresponding error bars.}
    \label{fig:dr_clv_comp}
\end{figure}

Since SB plus linear CLV was the best fit overall and the the introduction of CLV significantly reduces the contribution of the horizontal differential shear term $\alpha$, we want to make sure we are not confusing DR and CLV. Therefore, in Figure \ref{fig:dr_clv_comp} we compare the linear CLV from the SB fit and the differential shear contribution of the local RVs. Overall, we find the shape of the SB residuals matches closely with the linear CLV model and the curvature observed in the local RVs is a result of CLV. However, this is largely driven by the points which are at ingress (i.e. close to the stellar limb). Therefore, we cannot rule out the CLV signal detected could be caused by DR and more precision in the RVs would be needed in order to pick out DR plus a CLV contribution, especially if more observations are sampled at the limbs of the star.

\section{Discussion \& Conclusions}
\label{sec:con}
In this study, we have utilised two sets of NEID transit observations of the brown dwarf TOI-2119~b to determine the obliquity and conduct a study into stellar surface variability as a result of granulation. In addition, we used a number of photometric transit lightcurves from {\tess}, TMMT, and LCRO to update the system properties, including the ephemeris (see Table \ref{tab:observations}). To determine the obliquity we utilised the classical Rossiter McLaughlin and reloaded RM techniques. The RRM was also used to study stellar surface variability by determining the local velocities behind TOI-2119~b. For the classical RM we fit the disk integrated RVs with a model to determine a projected obliquity of $\lambda=-0.8\pm1.1~^\circ$ and a three-dimensional obliquity of $\psi=15.7_{-5.6}^{+5.4}~^\circ$. 

\begin{figure}
    \centering
    \includegraphics[width = 0.47\textwidth]{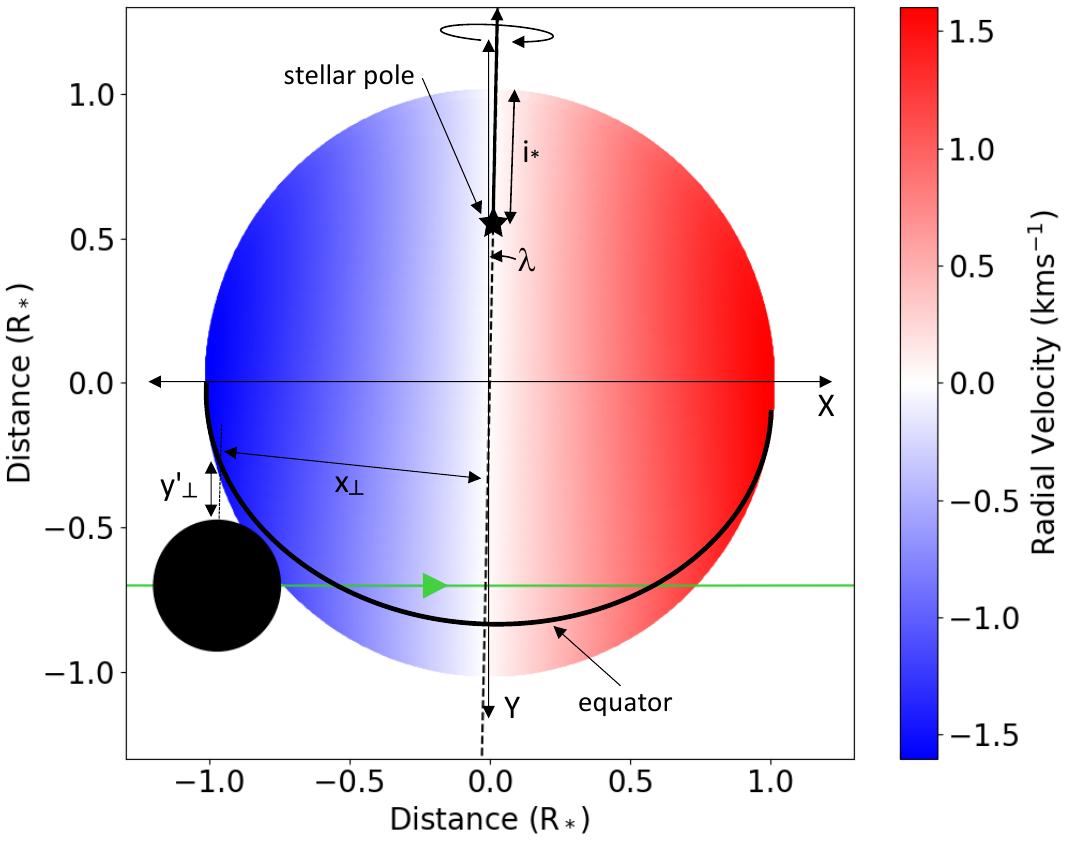}
    \caption{The projection of the TOI-2119 system in the plane of sky for the DR plus CLV linear model. The entire stellar disk is colour coded according to the radial velocity field. The stellar pole is indicated by a star and the stellar equator a solid black line, where the stellar inclination is assumed to be 33.2$^{\circ}$ from the DR towards plus CLV linear model, see Table \ref{tab:rrm_results}. The stellar spin axis is represented by a black line, solid when visible and dashed when hidden from view. The black circle represents TOI-2119~b to scale where its orbit is a green line.}
    \label{fig:schematic}
\end{figure}

\begin{figure}
    \centering
    \includegraphics[width = 0.47\textwidth]{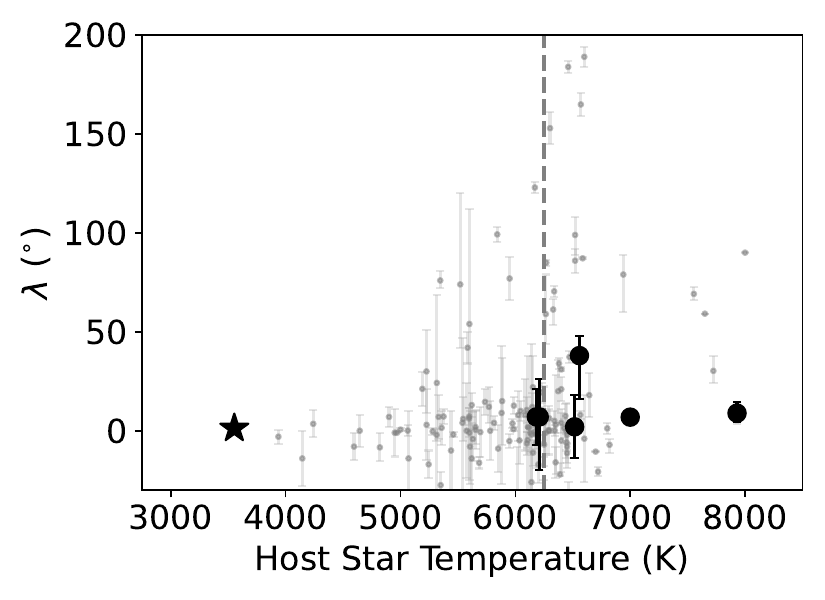}
    \caption{All known brown dwarf systems with measured obliquities as a function of the stellar temperature. The background grey sample are all gas giant planets (0.3~M$_{\rm{J}} \leq$ $M_{\rm{p}} \leq$ 13~$M_{\rm{J}}$) from TEPCat\protect\footnote{\url{https://www.astro.keele.ac.uk/jkt/tepcat/}{}}. The measured obliquity from the RRM method is shown for TOI-2119~b as the star marker. The obliquities of the other six known brown dwarf systems are: CoRoT-3~b \citep{triaud2009rossiter}, WASP-30~b \citep{triaud2013eblm}, KELT-1~b \citep{siverd2012kelt}, HATS-70~b \citep{zhou2019hats}, GPX-1~b \citep{giacalone2024oatmeal} and TOI-2533~b \citep{dos2024soles}. Note, at $\sim$6250~K both WASP-30~b and TOI-2533~b are plotted very close together.}
    \label{fig:BD_obl}
\end{figure}

Additionally, in the RRM technique we isolated the starlight behind the brown dwarf and fit local RVs with various models to account for stellar rotation (solid body and differential) and centre-to-limb convective variation (CLV) contributions, see Table \ref{tab:rrm_results}. With this method, our best fit model to the RVs is a solid body rotation case with a linear CLV contribution. This yields a projected obliquity of $\lambda = 1.26 \pm 1.3^{\circ}$, equatorial velocity of $v_{\rm{eq}}\sin i_\star = 1.61 \pm 0.1$~km s$^{-1}$ and linear CLV coefficient $c_1 = 1.09 \pm 0.2$. Similar to the classical RM, these values are consistent with TOI-2119~b being an aligned system in a prograde orbit where we adopt the values from the RRM technique to be the final results as CLV contributions are accounted for. To date, there are currently only six brown dwarf obliquities known, all of which are in aligned systems with obliquities $\lambda \leq$ 40.0$^{\circ}$, see Figure \ref{fig:BD_obl}. The classical RM in this work does not take into account differential rotation or centre-to-limb effects, therefore, when comparing only the SB RRM case the obliquities agree to within two sigma. Furthermore, the RVs used for the classical RM are derived from a pipeline on a limited wavelength range whereas the RRM analysis is conducted on CCFs which span the whole NEID wavelength range. This results in larger errors on the derived obliquity measurements from the RRM when compared to the classical RM obliquity. 

\footnotetext{\url{https://www.astro.keele.ac.uk/jkt/tepcat/}{}}

\begin{figure}
    \centering
    \subfloat{\includegraphics[width=0.47\textwidth]{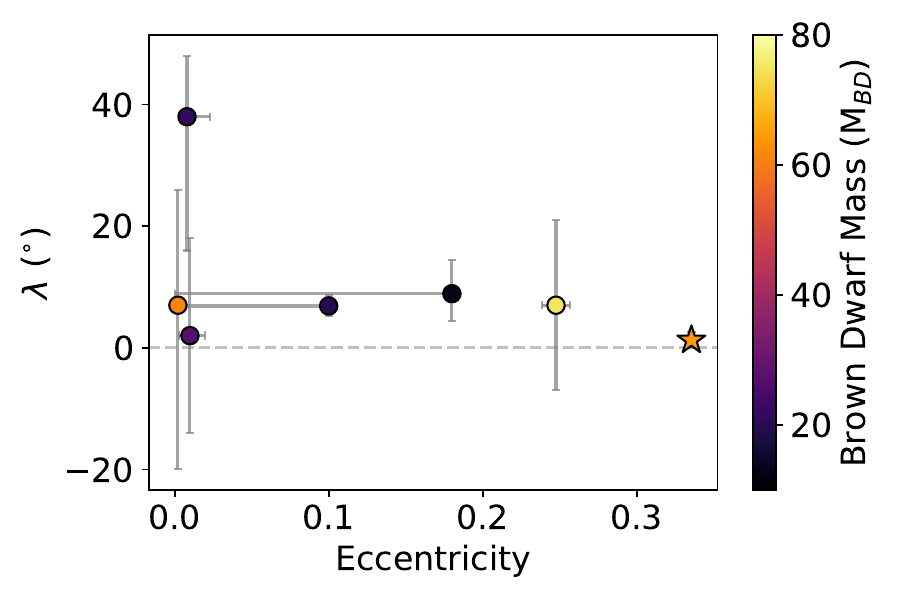}}\\ \vspace{-3mm}
    \subfloat{\includegraphics[width = 0.47\textwidth]{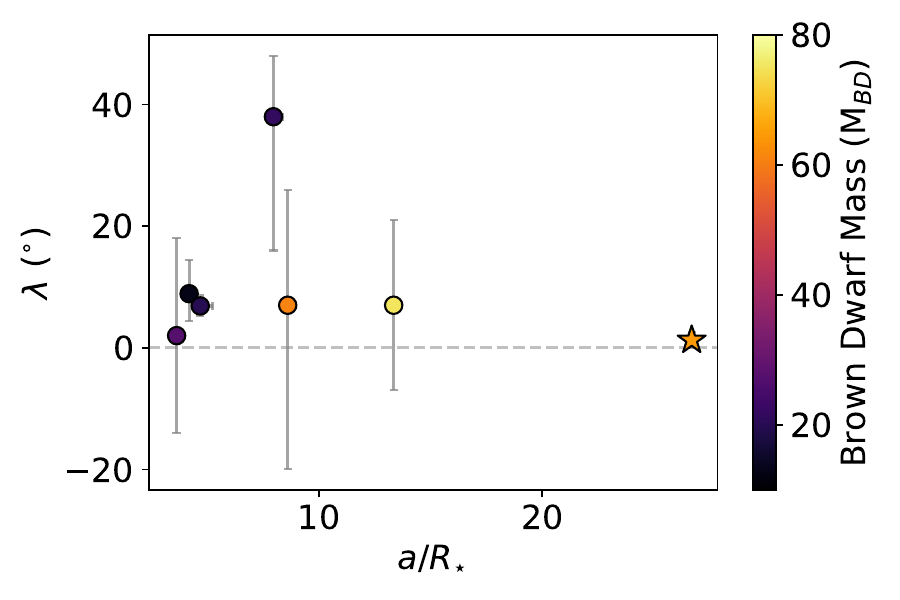}}
    \caption{The sky-projected obliquity known for all brown dwarf systems as a function of eccentricity ({\it Top}) and scaled semi-major axis, $a/R_{\star}$ ({\it Bottom}). Each point is colour coded according to the stellar host effective temperature. The star marker represents TOI-2119~b which has the highest eccentricity and scaled semi-major axis of all brown dwarfs with measured obliquities. A grey dashed line is added at $\lambda$ = 0$^{\circ}$. }
    \label{fig:BD_obl_ecc}
\end{figure}

Using the RRM we were able to explore the possibility of differential rotation (DR) of TOI-2119, finding a degeneracy between two scenarios either the star is pointing away from the observer or towards the observer; see Figure \ref{fig:schematic} for a schematic of the system. Using the stellar rotation period of 13.2~days and our $v_{\rm{eq}}\sin i_\star =$ 1.61~kms$^{-1}$, we can disentangle $i_\star$ from $v_{\rm{eq}}\sin i_\star$  yielding $i_\star\sim$55.4$^{\circ}$. From our DR models the best fit was the DR plus linear CLV for the towards the observer scenario with an $i_\star$ = 33.2$^{\circ}\substack{+34.4 \\ -19.3}$, which is in agreement with the derived $i_\star$ from the rotation period. For all DR models the differential shear $\alpha$ was determined to be consistent with zero, especially when CLV was taken into consideration. However, when we compared the effects of DR and CLV within the local RVs it was clear they followed the linear CLV model.

One of the most interesting questions related to brown dwarfs is how they formed and evolved, and if they form via the same mechanisms as hot Jupiters. Hot Jupiters (short orbital period Jupiter mass planets) are believed to form in a disc far from their stars and migrate to their observed orbits via disk migration or high-eccentricity tidal migration \cite{albrecht2022}. It has been argued the obliquity distribution for hot Jupiters favours high-eccentricity tidal migration as a result of the split at the Kraft break ($T_{\rm{eff}}$ = 6250~K) with hotter stars having higher obliquities than cooler stars \citep{winn2010hot,rice2022origins}. This is a direct result of the radiative envelopes of hotter stars being less efficient at tidally realigning stellar obliquities than the convective envelopes of cooler stars. Overall, the high obliquities of hot stars suggest that hot Jupiters migrate to their close-in orbits via a dynamically hot mechanism, such as high-eccentricity migration, rather than a dynamically cool mechanism, such as migration through the protoplanetary disk \citep[see][and references herein]{albrecht2022}.

High-eccentricity tidal migration involves a proto-planet being launched into an eccentric orbit as a result of either planet-planet scattering \citep{beauge2012multiple}, von Zeipel-Lidov-Kozai oscillations \citep{fabrycky2007shrinking}, or secular chaos \citep{wu2011secular}. Afterwards, tidal friction circularises and shrinks the planetary orbit over time. Therefore, amongst hot Jupiters, an association is expected between eccentricity and obliquity \citep{dawson2018origins}. The circularisation of an orbit is a much slower process when compared to the alignment of the stellar spin axis and synchronisation of stellar rotation rates, due to there being significantly more angular momentum in the orbit than in the stars \citep{hut1981tidal}. We calculated if TOI-2119~b could circularise in its lifetime using tidal circularisation timescales from \citet{bowler2020population} and assuming $Q_{\star} = Q_{\rm{BD}} = 10^{6}$. This resulted in a circularisation time much longer than the lifetime of TOI-2119~b  which is assumed to be between 0.7 - 5.1~Gyr \citep[see][based on scaling relationships from the 13.2~day rotation period]{canas2022eccentric}. Therefore, the system is too young for circularisation of the orbit to have occurred as a result of tidal effects during this time. However, \citet{rice2022origins} found cool stars ($T_{\rm{eff}}$ < 6100~K) with planets on eccentric orbits tend to have a higher obliquity, which is associated with a lower planet mass and a wide orbit. Therefore, for hot Jupiters and brown dwarfs orbiting a cool star it is possible there is no correlation between high obliquity and high eccentricity. From Figure \ref{fig:BD_obl}, it can be seen there are very few obliquity measurements for hot Jupiters around low mass stars with temperatures less than 4000~K. Similarly, hot Jupiters with measured obliquities around cool stars ($T_{\rm{eff}} \leq$ 5000~K) appear to be aligned which suggests a split in formation mechanism between hot and cool hosts.Additionally, \citet{hebrard2011transiting} and \citet{triaud2018rossiter} noted that objects more massive than $\sim$3~M$_{\rm{J}}$ tend to be in well aligned orbits which is seen amongst the brown dwarf population. Furthermore, In \citet{zhou2019hats} they highlight obliquities of all systems hosting 10~<~$M_p$~<~80~M$_{\rm{J}}$ companions, showing a lack of significantly misaligned systems within this mass range.

\citet{espinoza2023aligned} show there is an interesting number of aligned gas giant planets with eccentricity between 0.1 and 0.4. They also find this trend disappears for binary stars, for which there are three measurements WASP-8~b \citep{bourrier2017refined}, Kepler-420~b \citep{santerne2014sophie} and HD~80606~b \citep{hebrard2010observation}. Therefore, in Figure \ref{fig:BD_obl_ecc} we explore a similar trend in obliquity and eccentricity for all known brown dwarf systems with measured obliquities. We find these systems follow a similar trend as discovered in \citet{espinoza2023aligned} and find brown dwarf systems with eccentricity between 0.1 and 0.35 show an alignment in obliquity. However, we do note with only seven measurements it is a tentative trend which will only be strengthened with more obliquity measurements for brown dwarf systems. Due to the low obliquity and high eccentricity of the TOI-2119 system, it is unlikely high-eccentricity migration paths such as von Zeipel-Lidov-Kozai oscillations, planet-planet scattering, and secular chaos have played a role in the evolution of the system. However, another possible explanation is Coplanar High-eccentricity Migration \citep{petrovich2015hot}, where secular gravitational interactions between two planet/brown dwarfs in eccentric orbits with relatively low mutual inclinations and friction due to tides raised on the planet by the host star, allows for migration to occur on the same plane in which the planets/brown dwarfs formed. This scenario would require an as yet undetected and distant companion. However, TOI-2119 has a Renormalised Unit Weight Error (RUWE) of 1.93 in the Gaia DR3 \citep{vallenari2023gaiadr3} (this is not caused by the brown dwarf as other M dwarf/ brown dwarf systems have low RUWE values). RUWE is a goodness-of fit measurement of the single-star model to the targets astrometry which is highly sensitive to the photocentre motion of binaries \citep[see][]{belokurov2020unresolved}. Overall, the RUWE is expected to be 1.0 for single-star sources and greater than 1.4 for those which could be potential binary systems. The resulting value for TOI-2119 of 1.93 suggests that the star may in fact host a binary companion but further observations would be needed to confirm this. Direct imaging from NESSI \citep{scott2018nn} on the 3.5~m WYNN show no evidence of blending from a bright companion $\Delta$Mag < 4 at separations of 0\d{\arcsec}15–1\d{\arcsec}2 and further spectroscopic modelling show no stellar companions within 2$\arcsec$ \citep[see][for further details]{canas2022eccentric}. Additionally, RV constraints on additional companions were considered in the initial discovery paper of \citet{canas2022eccentric} where the presence of any additional low-inclination ($\sin i\sim$~1) brown dwarfs ($M_{\rm{BD}}$ < 11~M$_{\rm{J}}$) within 7.4 AU of TOI-2119. 

Approximately, 27\% of all known brown dwarfs orbit an M dwarf with some of these hosting an additional stellar companion \citep[e.g.][]{johnson2011lhs, irwin2010nltt, jackman2019ngts}. The spread in eccentricity for M dwarf/brown dwarf systems ranges from circular ($e$ = 0) to highly eccentric (i.e. $e >$ 0.7). For example TOI-1278~b has a mass of $M_{\rm{BD}}$ = 18.5~M$_{\rm{J}}$ and a low eccentricity of $e$ = 0.013. \citep{artigau2021toi}. On the other hand, NGTS-19~b has $M_{\rm{BD}}$ = 69.5~M$_{\rm{J}}$ and $e$ = 0.37 \citep{acton2021ngts}, while TOI-2490~b has $M_{\rm{BD}}$ = 73.6~M$_{\rm{J}}$ and $e$ = 0.78 \citep{henderson2024toi}. It appears there is a split in eccentricity between these systems (TOI-2119~b included) as a function of mass, see Figure \ref{fig:BD_obl_ecc} where scaled semi-major axis is also shown for completeness. \citet{bowler2020population} conduct an eccentricity study on giant planets ($\leq$ 15~M$_{\rm{J}}$) and brown dwarfs (15 - 75~M$_{\rm{J}}$) finding giant planets have lower eccentricities and brown dwarfs at all eccentricities. The simplest explanation for this is giant planets form in disks while brown dwarfs are forming at the lower mass end of star formation. Overall, it is clear there is a distinct difference in formation mechanism between giant planets, brown dwarfs and binary stars which is reliant on mass. For TOI-2119~b it is likely it formed as a result of star formation channels given its eccentricity, circulisation timescale, mass and age. As the number of brown dwarfs with obliquity measurements increases, this will form a complete picture of the relationship between obliquity and eccentricity and what they may tell us about brown dwarf formation and evolution. 

\section*{Acknowledgements}
This paper contains data taken with the NEID instrument (ID: 2023A-547291, PI: L. Doyle and ID: 2022A-802765, PI: C. Ca\~nas), which was funded by the NASA-NSF Exoplanet Observational Research (NN-EXPLORE) partnership and built by Pennsylvania State University. NEID is installed on the WIYN telescope, which is operated by the National Optical Astronomy Observatory, and the NEID archive is operated by the NASA Exoplanet Science Institute at the California Institute of Technology. NN-EXPLORE is managed by the Jet Propulsion Laboratory, California Institute of Technology under contract with the National Aeronautics and Space Administration. We also include data collected by the {\tess} mission, where funding for the {\tess} mission is provided by the NASA Explorer Program. 
Some of these results are based on observations obtained with the Apache Point Observatory 3.5m telescope, which is owned and operated by the Astrophysical Research Consortium. 

This research was funded in whole or in part by the UKRI, (Grants ST/X001121/1, EP/X027562/1). HMC acknowledges funding from a UKRI Future Leader Fellowship, grant number MR/S035214/1. 
C. Ca\~nas acknowledges support by NASA Headquarters through an appointment to the NASA Postdoctoral Program at the Goddard Space Flight Center, administered by USRA through a contract with NASA and the NASA Earth and Space Science Fellowship Program through grant 80NSSC18K1114. 
For the purpose of open access, the author has applied a Creative Commons Attribution (CC BY) licence (where permitted by UKRI, ‘Open Government Licence’ or ‘Creative Commons Attribution No-derivatives (CC BY-ND) licence’ may be stated instead) to any Author Accepted Manuscript version arising from this submission.

\section*{Data Availability}
The {\tess} data are available from the NASA MAST portal and the WIYN NEID data are public from the NEID data archive. The TMMT photometry is available through the discovery paper \citet{canas2022eccentric}. The remaining photometry (APO, LCRO etc.) is available on request from the authors.



\bibliographystyle{mnras}
\bibliography{TOI-2119}


\bsp	
\label{lastpage}
\end{document}